\documentclass[aps,nofootinbib,twocolumn,showpacs]{revtex4}
\usepackage{amsmath,amsfonts,amssymb}
\usepackage{enumitem}
\usepackage{graphicx}
\usepackage[caption=false]{subfig}

\begin{document}

\title{2 Fast 2 Fiducial:\\Gaussian processes for the interpolation and marginalization of waveform error\\in extreme-mass-ratio-inspiral parameter estimation}
\author{Alvin J. K. Chua}
\email{alvin.j.chua@jpl.nasa.gov}
\affiliation{Jet Propulsion Laboratory, California Institute of Technology, Pasadena, CA 91109, U.S.A.}
\author{Natalia Korsakova}
\email{natalia.korsakova@oca.eu}
\affiliation{Artemis, Observatoire de la C\^{o}te d'Azur, Boulevard de l'Observatoire, 06304 Nice, France}
\author{Christopher J. Moore}
\email{cmoore@star.sr.bham.ac.uk}
\affiliation{School of Physics and Astronomy, University of Birmingham, Edgbaston, Birmingham, B15 2TT, U.K.}
\author{Jonathan R. Gair}
\email{jonathan.gair@aei.mpg.de}
\affiliation{Max Planck Institute for Gravitational Physics, Potsdam Science Park, Am M\"{u}hlenberg 1, D-14476 Potsdam, Germany}
\author{Stanislav Babak}
\email{stas@apc.in2p3.fr}
\affiliation{Laboratoire Astroparticule et Cosmologie, 10 rue Alice Domon et L\'{e}onie Duquet, 75013 Paris, France}
\date{\today}

\begin{abstract}
A number of open problems hinder our present ability to extract scientific information from data that will be gathered by the near-future gravitational-wave mission LISA. Many of these relate to the modeling, detection and characterization of signals from binary inspirals with an extreme component-mass ratio of $\lesssim10^{-4}$. In this paper, we draw attention to the issue of systematic error in parameter estimation due to the use of fast but approximate waveform models; this is found to be relevant for extreme-mass-ratio inspirals even in the case of waveforms with $\gtrsim90\%$ overlap accuracy and moderate ($\gtrsim30$) signal-to-noise ratios. A scheme that uses Gaussian processes to interpolate and marginalize over waveform error is adapted and investigated as a possible precursor solution to this problem. Several new methodological results are obtained, and the viability of the technique is successfully demonstrated on a three-parameter example in the setting of the LISA Data Challenge. 
\end{abstract}

\pacs{02.50.-r, 04.80.Nn, 95.55.Ym, 95.75.Wx}
\maketitle

\section{Introduction}

In contrast to present ground-based gravitational-wave (GW) detectors, the future space interferometer LISA \cite{DEA2017} is expected to find an abundance of long-lived sources radiating in the millihertz band; among these are the extreme-mass-ratio inspirals (EMRIs) of compact objects (white dwarfs, neutron stars or stellar-mass black holes) into the supermassive black holes that reside at the centres of galaxies. Observations of EMRIs will complement electromagnetic astronomy in probing formation rates and evolution scenarios for supermassive black holes in galactic nuclei, and also provide measurements of strong-field gravity to unprecedented precision \cite{BEA2017,BEA2019}.

A typical EMRI will be observed by LISA for around $10^5$ orbits over the mission lifetime. Although this allows its properties to be measured very precisely from its gravitational waveform, the results are consequently susceptible to any inaccuracy in the efficiency-oriented waveform templates that are used in Bayesian inference algorithms. If this ``theoretical error'' is too large, it will dominate over statistical error at high or even moderate signal-to-noise ratios (SNRs) (as shown in \cite{CV2007}, for supermassive-black-hole mergers). Unfortunately, EMRIs are indeed difficult to model accurately, since the extreme mass ratio prohibits the use of fully numerical methods. The present state of the art is a perturbation-theory framework that computes the self-force effects of the compact object's gravitational field on its own orbit \cite{B2009,PPV2011}. Full waveform models that employ these calculations are still under development; they will also be computationally expensive, and are unlikely to be used directly for parameter estimation after they become available.

Current data analysis studies for LISA are therefore heavily reliant on the approximate EMRI template models known as kludges \cite{BC2004,BEA2007,CG2015,CMG2017,CGK2019}, which are designed for bulk use and may eventually be modified to include self-force information. However, it will still be challenging to do a coarse-grained detection search of the full EMRI parameter space with these templates \cite{GEA2004}, much less explore it with the precision required for parameter estimation. EMRI waveforms are extremely sensitive to small changes in their parameters, and so the global peak in the vast and multimodal posterior surface is akin to the proverbial needle in a haystack.\footnote{Or the not-so-proverbial bacterium on Earth \cite{WCW1998}, since $\sim10^{30}$ templates are required for full coverage of the parameter space.} This fact, coupled with the theoretical and computational difficulties of modeling the complex waveforms, makes EMRI search and inference the most formidable problem in LISA data analysis.

In this paper, we investigate the machine-learning technique of Gaussian-process regression (GPR) \cite{M2003,RW2006} as a possible strategy for mitigating theoretical error in EMRI parameter estimation. Specifically, GPR is used to interpolate a small set of precomputed waveform differences between a fiducial model and an approximate one; the GPR interpolant then provides a prior distribution for the waveform difference, which allows theoretical error to be marginalized over in the standard Bayesian likelihood with the approximate model \cite{MG2014,MBCG2016}. The benefits of this method for GW parameter estimation are twofold: it includes information from computationally expensive waveforms while searching with faster but less accurate templates, and also accounts for any residual model inaccuracy with more conservative error estimates.

An overview of the marginalized-likelihood method is given in Sec.~\ref{sec:marginalised_likelihood}. The technique of GPR (in the context of waveform interpolation) and the training of the Gaussian-process model from the precomputed set of waveform differences are introduced in Secs~\ref{subsec:GPR} and \ref{subsec:training}, respectively. Sec.~\ref{subsec:BBHs} then briefly summarizes results from a previous proof-of-concept study, where the method was applied to parameter estimation for comparable-mass black-hole binary mergers in the LIGO--Virgo--KAGRA sensitivity band \cite{MBCG2016}.

In Sec.~\ref{sec:EMRIs}, we investigate the viability of the method for EMRI parameter estimation, and deduce that the characteristic separation of points in the required training set is much greater than that in a notional set of posterior samples obtained with the fiducial waveforms. This is verified by heuristic one- and two-parameter examples in Secs~\ref{subsec:1D} and \ref{subsec:2D} respectively, where the (inverse) Fisher information is also shown to be a good substitute for the trained Gaussian-process covariance. We then apply our scheme in Sec.~\ref{subsec:LDC} to a modified, scaled-down version of the EMRI data set from the first round of the new LISA Data Challenge \cite{LDC2019}. Finally, possible computational strategies for generalizing the method to higher-dimensional searches are discussed in Sec.~\ref{sec:conclusion}.

\section{Marginalized likelihood}\label{sec:marginalised_likelihood}

In the standard matched-filtering framework for GW data analysis, single-source data from a two-channel detector can be written as the time series $x(t)=s(t)+n(t)$, where the source signal $s\equiv(s_\mathrm{I},s_\mathrm{II})$ is modeled as a deterministic function $h\equiv(h_\mathrm{I},h_\mathrm{II})$ of some astrophysical parameters $\boldsymbol{\lambda}$, while the detector noise $n\equiv(n_\mathrm{I},n_\mathrm{II})$ is assumed to be a Gaussian and stationary stochastic process. The Bayesian likelihood $L(\boldsymbol{\lambda})=p(x|\boldsymbol{\lambda})$ for the model parameters is defined as \cite{CF1994}
\begin{equation}\label{eq:standard_likelihood}
L\propto\exp{\left(-\frac{1}{2}\langle x-h|x-h\rangle\right)},
\end{equation}
with the noise-weighted inner product $\langle\cdot|\cdot\rangle$ on the space of finite-length time series given by
\begin{equation}\label{eq:inner_product}
\langle a|b\rangle=4\,\mathrm{Re}\sum_{f>0}^{f_N}\mathrm{d}f\sum_{\chi=\mathrm{I},\mathrm{II}}\frac{\tilde{a}_\chi^*(f)\tilde{b}_\chi(f)}{S_{n,\chi}(f)},
\end{equation}
where overtildes denote discrete Fourier transforms, $f_N$ is the Nyquist frequency, and $S_{n,\chi}(f)$ is the one-sided power spectral density of the channel noise $n_\chi$. A maximum-likelihood estimation $\boldsymbol{\lambda}_\mathrm{ML}$ of the parameters may then be obtained by maximising \eqref{eq:standard_likelihood} over the parameter space $\Lambda$, such that
\begin{equation}\label{eq:ML}
\left\langle\frac{\partial h}{\partial\boldsymbol{\lambda}}(\boldsymbol{\lambda}_\mathrm{ML})\big|x-h(\boldsymbol{\lambda}_\mathrm{ML})\right\rangle=\mathbf{0}.
\end{equation}

For a fiducial model $h_\mathrm{acc}$ that provides an accurate description of the source, the waveform at the true parameter values $\boldsymbol{\lambda}_\mathrm{true}$ corresponds to the source signal, i.e., $x=h_\mathrm{acc}(\boldsymbol{\lambda}_\mathrm{true})+n$. Any error $\boldsymbol{\lambda}_\epsilon=\boldsymbol{\lambda}_\mathrm{ML}-\boldsymbol{\lambda}_\mathrm{true}$ in the measured parameter values is then purely statistical, in that it is directly proportional to $n$. Using Einstein notation, we may write \eqref{eq:ML} at leading order in $\boldsymbol{\lambda}_\epsilon$ as
\begin{multline}\label{eq:statistical_error}
\langle[\partial h_\mathrm{acc}]_b|n-[\partial h_\mathrm{acc}]_a[\boldsymbol{\lambda}_\epsilon]^a\rangle\approx0\\
\implies[\boldsymbol{\lambda}_\epsilon]^a\approx[\Gamma_\mathrm{acc}^{-1}]^{ab}\langle n|[\partial h_\mathrm{acc}]_b\rangle,
\end{multline}
where the waveform derivative $\partial h$ and Fisher information matrix $\Gamma$ are defined respectively as
\begin{equation}\label{eq:derivatives}
[\partial h]_a=\frac{\partial h}{\partial[\boldsymbol{\lambda}]^a},\quad[\Gamma]_{ab}=\langle[\partial h]_a|[\partial h]_b\rangle,
\end{equation}
and evaluated at $\boldsymbol{\lambda}_\mathrm{ML}$.

In general, a template model $h_\mathrm{app}$ that is used for parameter estimation will only be approximate, such that $h_\mathrm{app}(\boldsymbol{\lambda}_\mathrm{true})\neq h_\mathrm{acc}(\boldsymbol{\lambda}_\mathrm{true})$. At leading order, any error in the measured parameter values may be written as
\begin{equation}\label{eq:theoretical_error}
[\boldsymbol{\lambda}_\epsilon]^a\approx[\Gamma_\mathrm{app}^{-1}]^{ab}\langle n|[\partial h_\mathrm{app}]_b\rangle-[\Gamma_\mathrm{app}^{-1}]^{ab}\langle h_\epsilon|[\partial h_\mathrm{app}]_b\rangle,
\end{equation}
where the first term is statistical in the sense of \eqref{eq:statistical_error}, and the second term corresponds to the theoretical error that arises from the difference $h_\epsilon$ between the approximate and accurate waveforms, i.e.,
\begin{equation}\label{eq:waveform_difference}
h_\epsilon=h_\mathrm{app}-h_\mathrm{acc}.
\end{equation}
Again, all derivatives in \eqref{eq:theoretical_error} are evaluated at $\boldsymbol{\lambda}_\mathrm{ML}$, while the waveform difference $h_\epsilon$ is evaluated at $\boldsymbol{\lambda}_\mathrm{true}$.\footnote{However, $h_\epsilon(\boldsymbol{\lambda}_\mathrm{true})=h_\epsilon(\boldsymbol{\lambda}_\mathrm{ML})$ at leading order, which allows the theoretical error in \eqref{eq:theoretical_error} (and $\boldsymbol{\lambda}_\mathrm{true}$ itself in the case of high SNR) to be estimated for a given measurement $\boldsymbol{\lambda}_\mathrm{ML}$ \cite{CV2007}.}

The statistical-error terms in \eqref{eq:statistical_error} and \eqref{eq:theoretical_error} are inversely proportional to the waveform SNR
\begin{equation}
\rho=\sqrt{\langle h|h\rangle},
\end{equation}
since $\partial h\propto\rho$ and $\Gamma\propto\rho^2$. On the other hand, the theoretical-error term in \eqref{eq:theoretical_error} is independent of $\rho$. Hence the systematic bias incurred by using approximate templates $h_\mathrm{app}$ in \eqref{eq:standard_likelihood} may dominate the noise uncertainty for high-SNR sources, and is likely to be the limiting factor in extracting parameter information from such signals \cite{CV2007}.

One approach to account for this bias is to marginalize over the error of $h_\mathrm{app}$ (with respect to $h_\mathrm{acc}$) in \eqref{eq:standard_likelihood}, by specifying a suitable prior probability distribution $p(h_\epsilon)$ for the waveform difference \cite{MG2014}. This ``marginalized likelihood'' is given by the (functional) integral
\begin{equation}\label{eq:marginalisation}
\mathcal{L}\propto\int_W\mathrm{D}h_\epsilon\,p(h_\epsilon)L_\mathrm{acc}
\end{equation}
on the space $W$ of waveform differences; it can be evaluated analytically if $p(h_\epsilon)$ is Gaussian, since $L_\mathrm{acc}$ (Eq.~\eqref{eq:standard_likelihood} with $h_\mathrm{acc}=h_\mathrm{app}-h_\epsilon$) is also formally Gaussian. Such a prior may be obtained through the technique of GPR, which provides an interpolant for $h_\epsilon$ with an associated (scalar) variance at each point in parameter space.

\subsection{Gaussian process regression}\label{subsec:GPR}

In the GPR approach, the waveform difference $h_\epsilon\in W$ is modeled as a zero-mean Gaussian process over $\Lambda$, i.e.,
\begin{equation}\label{eq:GP}
h_\epsilon\sim\mathcal{GP}(0,k),
\end{equation}
where the mean function is chosen (uninformatively) as the time series $0\in W$, and the covariance function $k(\boldsymbol{\lambda},\boldsymbol{\lambda}')$ is some symmetric and positive-definite bilinear form on $\Lambda$. For any finite set of parameter points $\{\boldsymbol{\lambda}_i\in\Lambda\,|\,i=1,2,...,N\}$, the corresponding set of waveform differences $\{h_\epsilon(\boldsymbol{\lambda}_i)\in W\,|\,i=1,2,...,N\}$ has a Gaussian probability distribution $\mathcal{N}(\mathbf{0},\mathbf{K})$ on $W^N$, i.e.,
\begin{equation}\label{eq:PDF}
p(h_\epsilon(\boldsymbol{\lambda}_i))\propto\frac{1}{\det{\mathbf{K}}}\exp{\left(-\frac{1}{2}\mathbf{v}^T\mathbf{K}^{-1}\mathbf{v}\right)},
\end{equation}
where the covariance matrix $\mathbf{K}$ and waveform difference vector $\mathbf{v}$ are given respectively by
\begin{equation}
[\mathbf{K}]_{ij}=k(\boldsymbol{\lambda}_i,\boldsymbol{\lambda}_j),
\end{equation}
\begin{equation}
[\mathbf{v}]_i=h_\epsilon(\boldsymbol{\lambda}_i).
\end{equation}

It is convenient to write the quadratic form in \eqref{eq:PDF} as
\begin{equation}
\mathbf{v}^T\mathbf{K}^{-1}\mathbf{v}=\mathrm{tr}\,(\mathbf{K}^{-1}\mathbf{M}),
\end{equation}
where
\begin{equation}\label{eq:overlap_matrix}
[\mathbf{M}]_{ij}=[\mathbf{v}\mathbf{v}^T]_{ij}=\frac{1}{\gamma}\langle h_\epsilon(\boldsymbol{\lambda}_i)|h_\epsilon(\boldsymbol{\lambda}_j)\rangle,
\end{equation}
with $\gamma>0$ the overall scale ratio between the frequency-averaged power spectral densities of the waveform differences and the detector noise. In choosing a frequency-independent form $k(\boldsymbol{\lambda},\boldsymbol{\lambda}')$ for the covariance function, we have assumed that the correlations among the waveform differences across parameter space do not depend on frequency. The waveform difference at each parameter point is also taken to be perfectly correlated across all frequency bins, which gives the particular normalizing factor in \eqref{eq:PDF}. Finally, the inner product for waveform differences in \eqref{eq:overlap_matrix} is chosen to be proportional to the noise-weighted one in \eqref{eq:inner_product}. These assumptions simplify the GPR calculations, but are also conservative in the sense that they generally yield less informative likelihoods; a more detailed justification is provided in \cite{MBCG2016}.

From the definition of a Gaussian process, the enlarged set $\{h_\epsilon(\boldsymbol{\lambda}_i),h_\epsilon(\boldsymbol{\lambda})\}$ is again normally distributed with zero mean and the covariance matrix
\begin{equation}
\mathbf{K}_*=\left[\begin{array}{cc}\mathbf{K} & \mathbf{k}_*\\\mathbf{k}_*^T & k_{**}\end{array}\right],
\end{equation}
where
\begin{equation}
[\mathbf{k}_*]_i=k(\boldsymbol{\lambda}_i,\boldsymbol{\lambda}),\quad k_{**}=k(\boldsymbol{\lambda},\boldsymbol{\lambda}).
\end{equation}
If $\{h_\epsilon(\boldsymbol{\lambda}_i)\}$ is known, then the conditional probability distribution of $h_\epsilon(\boldsymbol{\lambda})$ given $\{h_\epsilon(\boldsymbol{\lambda}_i)\}$ is also Gaussian, i.e.,
\begin{equation}\label{eq:conditional_PDF}
p(h_\epsilon(\boldsymbol{\lambda}))\propto\frac{1}{\sigma^2}\exp{\left(-\frac{1}{2}\frac{\langle h_\epsilon(\boldsymbol{\lambda})-\mu|h_\epsilon(\boldsymbol{\lambda})-\mu\rangle}{\gamma\sigma^2}\right)},
\end{equation}
where $\mu(\boldsymbol{\lambda})$ and $\sigma^2(\boldsymbol{\lambda})$ are given respectively by
\begin{equation}\label{eq:GPR_mean}
\mu=\mathbf{k}_*^T\mathbf{K}^{-1}\mathbf{v},
\end{equation}
\begin{equation}\label{eq:GPR_variance}
\sigma^2=k_{**}-\mathbf{k}_*^T\mathbf{K}^{-1}\mathbf{k}_*.
\end{equation}

The conditional probability \eqref{eq:conditional_PDF} forms the basis of GPR, and yields an interpolation of $h_\epsilon(\boldsymbol{\lambda})$ from a small, precomputed training set
\begin{equation}\label{eq:training_set}
\mathcal{D}=\{(\boldsymbol{\lambda}_i,h_\epsilon(\boldsymbol{\lambda}_i))\,|\,i=1,2,...,N\}.
\end{equation}
This interpolant is given by the waveform difference mean $\mu(\boldsymbol{\lambda})$, with associated variance $\sigma^2(\boldsymbol{\lambda})$; it essentially provides a new GPR-informed template model
\begin{equation}\label{eq:GPR_waveform}
h_\mathrm{GPR}=h_\mathrm{app}-\mu,
\end{equation}
which approximates $h_\mathrm{acc}$ via \eqref{eq:waveform_difference}. Eq.~\eqref{eq:conditional_PDF} also supplies the prior for $h_\epsilon$ in \eqref{eq:marginalisation}, which evaluates to
\begin{equation}\label{eq:marginalised_likelihood}
\mathcal{L}\propto\frac{1}{1+\gamma\sigma^2}\exp{\left(-\frac{1}{2}\frac{\langle x-h_\mathrm{GPR}|x-h_\mathrm{GPR}\rangle}{1+\gamma\sigma^2}\right)}.
\end{equation}

The GPR marginalized likelihood has several desirable features for parameter estimation. A maximum-likelihood estimation of $\boldsymbol{\lambda}$ with \eqref{eq:marginalised_likelihood} gives
\begin{equation}
\langle[\partial h_\mathrm{GPR}]_b|n-h_\epsilon+\mu-[\partial h_\mathrm{GPR}]_a[\boldsymbol{\lambda}_\epsilon]^a\rangle\approx0,
\end{equation}
from \eqref{eq:ML} with $x=h_\mathrm{acc}(\boldsymbol{\lambda}_\mathrm{true})+n$ and $\boldsymbol{\lambda}_\epsilon=\boldsymbol{\lambda}_\mathrm{ML}-\boldsymbol{\lambda}_\mathrm{true}$. Hence we have
\begin{multline}
[\boldsymbol{\lambda}_\epsilon]^a\approx[\Gamma_\mathrm{GPR}^{-1}]^{ab}\langle n|[\partial h_\mathrm{GPR}]_b\rangle-[\Gamma_\mathrm{GPR}^{-1}]^{ab}\langle h_\epsilon|[\partial h_\mathrm{GPR}]_b\rangle\\+[\Gamma_\mathrm{GPR}^{-1}]^{ab}\langle\mu|[\partial h_\mathrm{GPR}]_b\rangle,
\end{multline}
where the third term is proportional to the GPR interpolant $\mu$, and acts to cancel the second term by design. This correction greatly reduces the systematic bias due to theoretical error, provided the interpolant is performing optimally (i.e., $\mu\approx h_\epsilon$) near $\boldsymbol{\lambda}_\mathrm{true}$.

Another safeguard against theoretical error is the presence of the GPR variance $\sigma^2$ in \eqref{eq:marginalised_likelihood}. This variance is $\ll1$ when $\mu\approx h_\epsilon$, but may become $\sim1$ far from all training-set points, or in the case of a suboptimally chosen training set or covariance function. The density in \eqref{eq:marginalised_likelihood} is then typically (but not necessarily) broadened over the accurate likelihood $L_\mathrm{acc}$, which is conservative as it acts to prevent the true parameter values from being excluded at high significance.

Lastly, the premise of the GPR approach is based on the availability of fiducial waveforms $h_\mathrm{acc}$ that are extremely expensive to compute, and hence unsuitable for use in Monte Carlo search algorithms with the standard likelihood. The marginalized likelihood remains computationally tractable while including information from $h_\mathrm{acc}$, since it only uses the approximate templates $h_\mathrm{app}$, and adds to them some linear combination of precomputed waveform differences via \eqref{eq:GPR_mean}. Any extra computational cost from using the marginalized likelihood thus scales linearly with the size $N$ of the training set. The scaling coefficient (relative to the cost of \eqref{eq:standard_likelihood}) is typically small; for the analyses in Sec.~\ref{sec:EMRIs}, it is $\sim10^{-3}$.

\subsection{Training the Gaussian process}\label{subsec:training}

With the zero-mean assumption in \eqref{eq:GP}, the waveform difference model is fully specified by the covariance function $k$. The standard approach is to define a functional form for $k$ that depends on a number of hyperparameters $\boldsymbol{\theta}$, and to select values for $\boldsymbol{\theta}$ by training the Gaussian process with information from the set $\mathcal{D}$. A covariance function $k(\boldsymbol{\lambda},\boldsymbol{\lambda}')$ is stationary if it depends only on the relative position $\boldsymbol{\lambda}-\boldsymbol{\lambda}'$ of the two parameter points; it is furthermore isotropic if it depends only on
\begin{equation}\label{eq:parameter_distance}
\tau^2=g_{ab}[\boldsymbol{\lambda}-\boldsymbol{\lambda}']^a[\boldsymbol{\lambda}-\boldsymbol{\lambda}']^b,
\end{equation}
where the $g_{ab}$ are the $\ell(\ell+1)/2$ independent components of some constant parameter-space metric $\mathbf{g}$ on $\Lambda$ (with $\ell=\dim{(\Lambda)}$).

An investigation of various common isotropic (hence stationary) covariance functions in the GW context finds the performance of the GPR interpolant and the marginalized likelihood to be fairly robust against changes in the functional form for $k$ \cite{MBCG2016}. Hence we consider a single fixed form in this paper: the squared-exponential covariance function
\begin{equation}\label{eq:SE}
k_\mathrm{SE}(\boldsymbol{\lambda},\boldsymbol{\lambda}')=\sigma_f^2\exp{\left(-\frac{1}{2}\tau^2\right)},
\end{equation}
which is the smooth limiting case for several different families of covariance functions. The hyperparameters for the model $\mathcal{GP}(0,k_\mathrm{SE})$ then comprise only the metric components $g_{ab}$ and some overall scale factor $\sigma_f^2$.

As the size of the training set $\mathcal{D}$ increases, the covariance matrix $\mathbf{K}$ rapidly becomes ill-conditioned, even for a modestly sized set with $N\gtrsim10$. This is partly mitigated by the addition of noise to $\mathcal{D}$, such that the GPR interpolant need only pass close to---rather than through---each training-set point. We transform
\begin{equation}\label{eq:noise}
[\mathbf{K}]_{ij}\to[\mathbf{K}]_{ij}+\sigma_f^2\sigma_n^2\delta_{ij},
\end{equation}
where $\delta_{ij}$ is the Kronecker delta, and the fractional noise variance $\sigma_n^2$ of each training-set point is taken to be uniform and fixed (i.e., not treated as a hyperparameter). In practical terms, the transformation \eqref{eq:noise} effectively reduces the condition number of $\mathbf{K}$, thereby facilitating its numerical inversion. We use $\sigma_n^2=10^{-4}$ throughout this paper, which is the smallest value compatible with all of the $N\lesssim100$ training sets considered in Sec.~\ref{sec:EMRIs}.

The most straightforward method of selecting the Gaussian-process hyperparameters $\boldsymbol{\theta}\equiv(g_{ab},\sigma_f^2)$ is through maximum-likelihood estimation with the hyperlikelihood $Z(\boldsymbol{\theta}|\mathcal{D})=p(h_\epsilon(\boldsymbol{\lambda}_i))$ from \eqref{eq:PDF}, i.e., the likelihood for the hyperparameters given the training set. In other words, an optimal set of hyperparameters $\boldsymbol{\theta}_\mathrm{ML}$ is obtained by maximizing the log-hyperlikelihood
\begin{equation}\label{eq:log-hyperlikelihood}
\ln{Z}=-\frac{1}{2}\mathrm{tr}\,(\mathbf{K}^{-1}\mathbf{M})-\ln{\det{\mathbf{K}}}+\mathrm{const}
\end{equation}
over the hyperparameter space $\Theta$.

Part of this maximization may be done analytically, since the overall scale $\sigma_f^2$ factors out of the matrix expressions in $\ln{Z}$ \cite{MCBG2016}. In the case of \eqref{eq:log-hyperlikelihood}, $\ln{Z}$ with $\mathbf{K}=\sigma_f^2\hat{\mathbf{K}}$ achieves a maximum in $\sigma_f^2$ at
\begin{equation}\label{eq:sigma_f}
\sigma_f^2=\frac{1}{2N}\mathrm{tr}\,(\hat{\mathbf{K}}^{-1}\mathbf{M}).
\end{equation}
Note that $\sigma_f^2$ contains a factor of $1/\gamma$ through \eqref{eq:overlap_matrix}, which effectively cancels out the appearance of $\gamma$ in \eqref{eq:conditional_PDF} and \eqref{eq:marginalised_likelihood}.\footnote{This may not be desirable in practice, since we may want to preserve the overall scale ratio between the waveform-difference and noise power spectral densities by fixing $\sigma_f^2$; see Sec.~\ref{subsec:LDC}.} Substituting \eqref{eq:sigma_f} back into \eqref{eq:log-hyperlikelihood} then gives a scale-invariant form for $\ln{Z}$, i.e.,
\begin{equation}\label{eq:scale-invariant}
\ln{Z}=-N\ln{\mathrm{tr}\,(\mathbf{K}^{-1}\mathbf{M})}-\ln{\det{\mathbf{K}}}+\mathrm{const}.
\end{equation}
Eqs~\eqref{eq:sigma_f} and \eqref{eq:scale-invariant} effectively reduce the dimensionality of the hyperparameter space by one (to $\dim{(\Theta)}=\ell(\ell+1)/2$ for the model $\mathcal{GP}(0,k_\mathrm{SE})$), which is useful for the low-dimensional searches conducted in Sec.~\ref{sec:EMRIs}.

\subsection{Previous application to binary black holes}\label{subsec:BBHs}

The viability of the GPR marginalized likelihood for improving GW parameter estimation has previously been demonstrated through a one-parameter ($\ell=1$) study, using waveforms for merging black-hole binary systems with comparable component masses $(m_1,m_2)$ \cite{MBCG2016}. Two waveform models implemented in the LIGO Scientific Collaboration Algorithm Library \cite{LSC2016} were considered: the phenomenologically fitted IMRPhenomC \cite{SEA2010} and the analytic TaylorF2 \cite{DIS2001}, which were taken as accurate and approximate respectively. Even though these two waveforms are qualitatively different (IMRPhenomC describes the full inspiral--merger--ringdown while TaylorF2 is inspiral-only), the marginalized likelihood functioned as described in reducing systematic bias.

In \cite{MBCG2016}, the marginalized likelihood was used to estimate the chirp mass $\mathcal{M}=(m_1m_2)^{3/5}/(m_1+m_2)^{1/5}$ from synthetic data $x=h_\mathrm{acc}(\boldsymbol{\lambda}_\mathrm{true})$ (with a zero realization of detector noise for simplicity), where $h_\mathrm{acc}(\boldsymbol{\lambda}_\mathrm{true})$ was an injected IMRPhenomC signal with $\mathcal{M}_\mathrm{true}=5.045M_\odot$ and fixed mass ratio $m_1/m_2=0.75$. As the density of the training set (with respect to some metric on $\Lambda$) was expected to be the strongest determinant of interpolation performance, two different grid lengths in $\mathcal{M}$ were considered: $\Delta\mathcal{M}=10^{-2}M_\odot$ and $\Delta\mathcal{M}=5\times10^{-3}M_\odot$. The training-set points were gridded uniformly around $\mathcal{M}_\mathrm{true}$ across the range $5\leq\mathcal{M}/M_\odot\leq5.6$, such that the density of the set was varied by fixing its span and changing its cardinality. A GPR model with the squared-exponential covariance function \eqref{eq:SE} was trained on both sets by optimizing the single independent hyperparameter, which was chosen more intuitively in this $\ell=1$ case to be the covariance length $\delta\mathcal{M}=(g_{\mathcal{M}\mathcal{M}})^{-1/2}$. In general, the optimal value of $\delta\mathcal{M}$ was found to change with the density and cardinality of the training set, but typically by less than a factor of two. The performance of the marginalized likelihood was found to be similarly robust against the choice of $\delta\mathcal{M}$ for a given training set.

Unsurprisingly, the performance of the marginalized likelihood was improved for the denser training set. Higher fidelity between the GPR waveform \eqref{eq:GPR_waveform} and the accurate waveform was obtained across the span of the set; this was quantified by overlaps $\mathcal{O}(h_\mathrm{GPR}|h_\mathrm{acc})$ that were $\gtrsim0.999$, with $\mathcal{O}(\cdot|\cdot)$ defined as
\begin{equation}
\mathcal{O}(a|b)=\frac{\langle a|b\rangle}{\sqrt{\langle a|a\rangle\langle b|b\rangle}}.
\end{equation}
The variance $\sigma^2$ associated with the waveform difference interpolant was smaller for the denser training set as well, with values that were $\lesssim10^{-3}$ relative to $\sigma_f^2$ (the limiting value of $\sigma^2$ outside the span of the set). A maximum-likelihood estimation of $\mathcal{M}$ with the corresponding marginalized likelihood was therefore closer to the true value, and better constrained. The sparser training set gave $\mathcal{O}(h_\mathrm{GPR}|h_\mathrm{acc})\gtrsim0.985$ and $\sigma^2/\sigma_f^2\lesssim10^{-2}$, with a marginalized likelihood that was discernibly worse but still functional. As will be discussed in Sec.~\ref{sec:EMRIs}, this is because its density was close to some threshold determined by the optimal value of $\delta\mathcal{M}$ (which is largely independent of training-set density).

Different source SNRs in the range $8\leq\rho\leq64$ were also considered in \cite{MBCG2016}. The marginalized likelihood for the sparser training set reduced the systematic error in the maximum-likelihood estimation of $\mathcal{M}$ from $\mathcal{M}_\epsilon=5\times10^{-3}M_\odot$ (around 10-sigma for a typical LIGO source with $\rho=16$) to $\mathcal{M}_\epsilon=9\times10^{-4}M_\odot$; it also broadened to remain consistent with $\mathcal{M}_\mathrm{true}$ at 2-sigma, even at high SNR. These results were obtained in the regime where the overlaps between the accurate and approximate waveforms were $\approx0.35$ across the span of the training set. Although theoretical error will be reduced if the approximate model is improved, results in Sec.~\ref{sec:EMRIs} show that the marginalized likelihood remains relevant for overlaps as high as $0.97$, which may still lead to significant systematic bias for a typical EMRI with $\rho=30$.

\section{Application to extreme-mass-ratio inspirals}\label{sec:EMRIs}

The detection and characterization of EMRIs is a formidable challenge in GW data analysis, especially in the broader context of resolving these sources from a LISA data set that is likely to contain an (over)abundance of long-lived and overlapping signals \cite{RC2017}. Even as a standalone problem, searches of the EMRI parameter space are greatly hindered by its large volume as measured by the Fisher information metric, which suggests $\sim10^{30}$ waveforms are required for full coverage in a template bank approach \cite{GEA2004}. This is exacerbated by the long and unwieldy templates themselves; a sampling rate of 0.2\,Hz (the approximate Nyquist rate for an EMRI with a $10^6M_\odot$ central black hole) yields $\sim10^7$ samples for each channel of a year-long signal.

Due to the $O(N^3)$ cost of computing \eqref{eq:GPR_mean} and \eqref{eq:GPR_variance}, it is clearly impractical---if not impossible---to cover any significant fraction of parameter space with a single training set. The present purpose of the GPR marginalized likelihood is thus restricted to precise parameter estimation in highly localized regions of parameter space. Furthermore, if the GPR approach is to be useful for EMRI inference at all, the typical separation of points in the training set must be significantly greater than the Fisher metric lengths of the fiducial waveform model \cite{C2016}, which determine the sampling density required to resolve and reconstruct the Bayesian posterior (otherwise generating the training set would be as expensive as directly sampling with the accurate waveform, which is intractable).

A simple argument shows that this is normally the case for waveforms (or waveform differences) $h(\boldsymbol{\lambda})$ with $\rho>1$. We consider a small neighbourhood of some point $\boldsymbol{\lambda}_0\in\Lambda$, along with a covariance metric $g_{ab}$ for a Gaussian process that accurately describes the distribution of $h$ over that neighbourhood. The metric defines the short covariance lengths $[\delta\boldsymbol{\lambda}]^a=(g_{aa})^{-1/2}$, i.e., the half-widths of the associated hyperellipsoid when restricted to each one-dimensional parameter subspace through $\boldsymbol{\lambda}_0$. These lengths place upper bounds on the characteristic grid lengths of the training set and lower bounds on its span, since a training set with larger grid lengths or a smaller span typically yields no peak in the log-hyperlikelihood surface \eqref{eq:log-hyperlikelihood}, and so the regression becomes suboptimal. Nevertheless, we may always choose grid lengths as large as $\sim [\delta\boldsymbol{\lambda}]^a$ if required.

From the assumption that the Gaussian process describes the distribution of $h$ accurately, the optimal covariance lengths $\delta\boldsymbol{\lambda}$ approximate the correlation lengths of $h$ itself. Hence we have $[\delta\boldsymbol{\lambda}]^a\sim[\delta\boldsymbol{\lambda}_{\mathrm{over}}]^a$, where the overlap lengths $\delta\boldsymbol{\lambda}_\mathrm{over}$ are defined to satisfy
\begin{equation}
\langle h(\boldsymbol{\lambda}_0)|h(\boldsymbol{\lambda}_0+P_a\delta\boldsymbol{\lambda}_\mathrm{over})\rangle=0
\end{equation}
for each $a$, with $P_a$ projecting $\delta\boldsymbol{\lambda}_\mathrm{over}$ onto the subspace corresponding to $[\boldsymbol{\lambda}]^a$. At leading order, we then have
\begin{equation}\label{eq:covariance_lengths}
[\delta\boldsymbol{\lambda}]^a\sim\frac{\langle h|h\rangle}{|\langle h|[\partial h]_a\rangle|}=\frac{\rho|\sec{[\boldsymbol{\phi}]^a}|}{\sqrt{[\Gamma]_{aa}}},
\end{equation}
where $\partial h$ and $\Gamma$ are defined as in \eqref{eq:derivatives}, and $[\boldsymbol{\phi}]^a$ is the principal inner-product angle between $h$ and $[\partial h]_a$.

Since the short (i.e., defined analogously to $[\delta\boldsymbol{\lambda}]^a$) Fisher metric lengths are given by
\begin{equation}\label{eq:Fisher_lengths}
[\delta\boldsymbol{\lambda}_\mathrm{Fish}]^a=\frac{1}{\sqrt{[\Gamma]_{aa}}},
\end{equation}
it follows that $[\delta\boldsymbol{\lambda}]^a\gtrsim\rho[\delta\boldsymbol{\lambda}_\mathrm{Fish}]^a$, as required for the argument. In general, any waveform derivative with respect to a parameter that only affects amplitude will give $[\boldsymbol{\phi}]^a\approx0$, such that we have $[\delta\boldsymbol{\lambda}]^a\sim\rho[\delta\boldsymbol{\lambda}_\mathrm{Fish}]^a$. Even for such parameters, it is still possible to interpolate $\rho>1$ waveforms with a training set whose density is below that of a typical set of posterior samples.

Eqs~\eqref{eq:covariance_lengths} and \eqref{eq:Fisher_lengths} show that the optimal covariance lengths (hence permissible grid lengths) are largely independent of SNR while the Fisher metric lengths are $O(1/\rho)$, and so the computational benefits of using the GPR marginalized likelihood over the standard likelihood with accurate waveforms are enhanced for sources with higher SNR. Furthermore, the Fisher lengths (rescaled by SNR) give reasonable estimates of the optimal covariance lengths, and are more straightforward to obtain. We show in the following sections that the Fisher matrix can provide good initial guesses for the covariance metric when maximizing the log-hyperlikelihood with standard optimization routines, or even serve as a proxy for the metric itself (i.e., foregoing the actual training procedure altogether). Consequently, it may also be used to specify the placement of the precomputed waveform differences.

As pointed out in \cite{MBCG2016}, the Fisher metric lengths of the difference between two waveform models are generally larger than those of the individual models, especially if both models generate waveforms with high overlap. However, the above argument implies it is actually the Fisher metric for the normalized (i.e., unit-SNR) waveform difference that is relevant to the GPR approach. Furthermore, for the examples in the following sections, this is found to be approximately proportional to the Fisher metric for both the accurate and approximate unit-SNR waveforms (which are comparable themselves), with a proportionality factor of $\sim1$. In other words, the waveform difference varies across parameter space in similar fashion to both waveforms. We make use of this observation in Sec.~\ref{subsec:LDC}, where the numerical derivatives of the accurate waveform, and hence the waveform-difference Fisher matrix, are expensive to compute correctly.

The validity of the above argument---and the viability of the GPR marginalized likelihood for EMRIs---is illustrated through heuristic one- and two-parameter analyses in Secs~\ref{subsec:1D} and \ref{subsec:2D} respectively; the scheme is then put into practice on a more realistic data set in Sec.~\ref{subsec:LDC}, which describes the noisy time-delay-interferometry (TDI) response \cite{AET1999} of the LISA instrument to an isolated EMRI signal. Due to their computational practicality, a variety of kludges (mixed-formalism EMRI template models) with different degrees of accuracy are used throughout this study, as either the fiducial or approximate waveform. The implementations of these are publicly available as part of the EMRI Kludge Suite (\texttt{github.com/alvincjk/EMRI\_Kludge\_Suite}) \cite{CGK2019}.

\subsection{1-D parameter estimation}\label{subsec:1D}

The fiducial model in this section (and Sec.~\ref{subsec:2D}) is taken to be the semi-relativistic numerical kludge (NK) \cite{BEA2007}, which has high fidelity with the more accurate Teukolsky-based models \cite{DH2006} up to an orbital separation of $\approx5M$. A recent augmentation \cite{CG2015,CMG2017,CGK2019} of the analytic kludge \cite{BC2004} is used as the approximate model; this augmented analytic kludge (AAK) is faster than the NK and matches its early phase evolution with high waveform overlap, but dephases gradually as the compact object approaches plunge. We take the data $x$ to be an NK signal from a $10^1M_\odot$ stellar-mass black hole orbiting a $10^6M_\odot$ central black hole, in the long-wavelength approximation \cite{C1998} and with a zero realization of LISA noise. The signal is two months long and sampled at 0.2\,Hz, while the source distance is adjusted to specify an SNR of $\rho=30$. Other orbital parameters are chosen such that the NK and AAK waveforms (for the same injected parameter values) have an overlap of 0.97, so as to investigate the scenario in which the approximate model is fairly accurate to begin with.

In this section, the GPR marginalized likelihood \eqref{eq:marginalised_likelihood} is used to estimate only the compact-object mass $\mu_\mathrm{true}=10^1M_\odot$, assuming all other parameters are known and fixed at their true values. The covariance metric on the corresponding parameter subspace has a single component $g_{\mu\mu}$, which is optimized through the maximization of the log-hyperlikelihood \eqref{eq:log-hyperlikelihood}. This one-dimensional example provides a simple illustration of the relationships between the optimal covariance length $\delta\mu=(g_{\mu\mu})^{-1/2}$, the Fisher metric length $\delta\mu_\mathrm{Fish}=([\Gamma]_{\mu\mu})^{-1/2}$, and the training-set grid length $\Delta\mu$.

The GPR model is trained on eight 10-point training sets with uniform grids, where the grid lengths are distributed in the range $-2.1\leq\lg{(\Delta\mu/M_\odot)}\leq-1.4$. This range is chosen to encompass the Fisher length of the unit-SNR waveform difference, which is approximately constant across the spans of the considered training sets and evaluated at $\mu_\mathrm{true}$ as $\lg{(\delta\mu_\mathrm{Fish}/M_\odot)}=-1.96$.\footnote{For comparison, the Fisher lengths of the unit-SNR NK and AAK waveforms at $\mu_\mathrm{true}$ are $\lg{(\delta\mu_\mathrm{Fish}/M_\odot)}=-1.86$ and $\lg{(\delta\mu_\mathrm{Fish}/M_\odot)}=-1.85$ respectively.} Each training set is placed such that $\mu_\mathrm{true}$ lies at the geometric centre of its span, and thus maximally far from the nearest points in the set.

\begin{figure}
\centering
\includegraphics[width=\columnwidth]{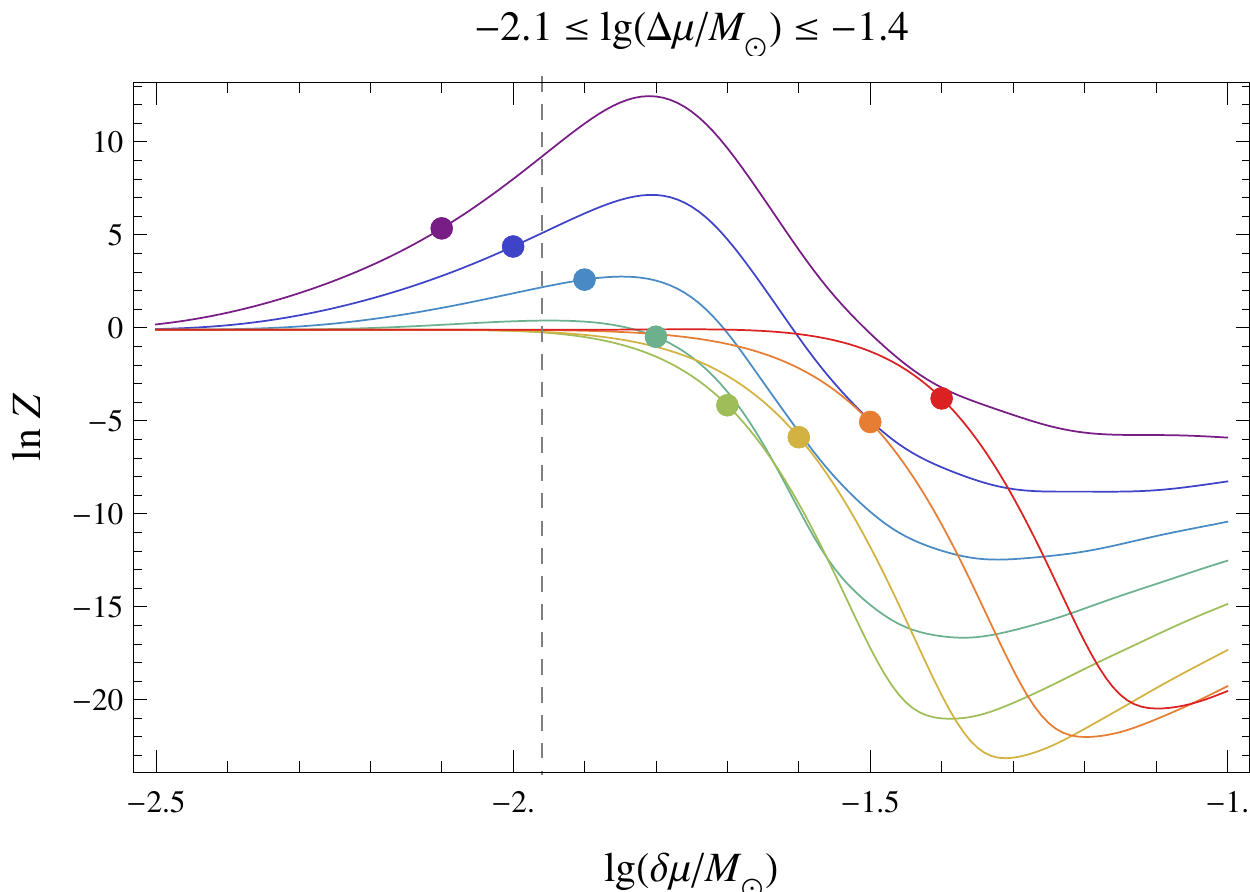}
\caption{Plots of $\ln{Z}$ against $\lg{(\delta\mu/M_\odot)}$ for eight 10-point training sets with grid lengths $-2.1\leq\lg{(\Delta\mu/M_\odot)}\leq-1.4$ (indicated by the abscissae of the solid circles). The vertical dashed line corresponds to the Fisher metric length $\delta\mu_\mathrm{Fish}$.}
\label{fig:1D_hyperlikelihood}
\end{figure}

Fig.~\ref{fig:1D_hyperlikelihood} shows plots of the log-hyperlikelihood for the eight training sets, with the optimal covariance length for each set given by the abscissa of the peak (where it exists). The optimal value $\delta\mu$ is approximately constant for all valid training sets, and falls in the narrow range $\lg{(\delta\mu_\mathrm{Fish}/M_\odot)}\leq\lg{(\delta\mu/M_\odot)}\leq-1.8$. In comparison to the approach of \cite{MBCG2016} described in Sec.~\ref{subsec:BBHs}, varying the density of the training set here by fixing its cardinality and changing its span also shifts $\delta\mu$ by less than a factor of two, which implies that both span and cardinality have less impact than density on a training set's performance. Hyperlikelihood peaks emerge only for grid lengths $\lg{(\Delta\mu/M_\odot)}\leq-1.8$, indicating that $1/\delta\mu$ corresponds approximately to a minimum threshold for the density of an optimal training set. Finally, $\delta\mu_\mathrm{Fish}$ appears to set a lower bound on $\delta\mu$, which is consistent with the discussion around \eqref{eq:covariance_lengths} and \eqref{eq:Fisher_lengths}.

We now consider the marginalized likelihood itself for three other 10-point training sets. Firstly, a set $\mathcal{D}_\mathrm{Fish}$ is placed around $\mu_\mathrm{true}$ with grid length $\Delta\mu_\mathrm{Fish}=\delta\mu_\mathrm{Fish}$; training the GPR model on this set yields an optimal covariance length $\lg{(\delta\mu/M_\odot)}=-1.82$. Two more training sets $\mathcal{D}_\mathrm{cov}$ and $\mathcal{D}_\mathrm{2cov}$ are constructed in the same way, with the grid lengths $\Delta\mu_\mathrm{cov}=\delta\mu$ and $\Delta\mu_\mathrm{2cov}=2\delta\mu$ respectively. A different optimal covariance length is found for $\mathcal{D}_\mathrm{cov}$, while there is no hyperlikelihood peak for $\mathcal{D}_\mathrm{2cov}$. Nevertheless, the above value of $\delta\mu$ is used for all three training sets, as the performance of the marginalized likelihood with each set is found to be practically constant across the range $\lg{(\delta\mu_\mathrm{Fish}/M_\odot)}\leq\lg{(\delta\mu/M_\odot)}\leq-1.8$.

\begin{figure}
\centering
\includegraphics[width=\columnwidth]{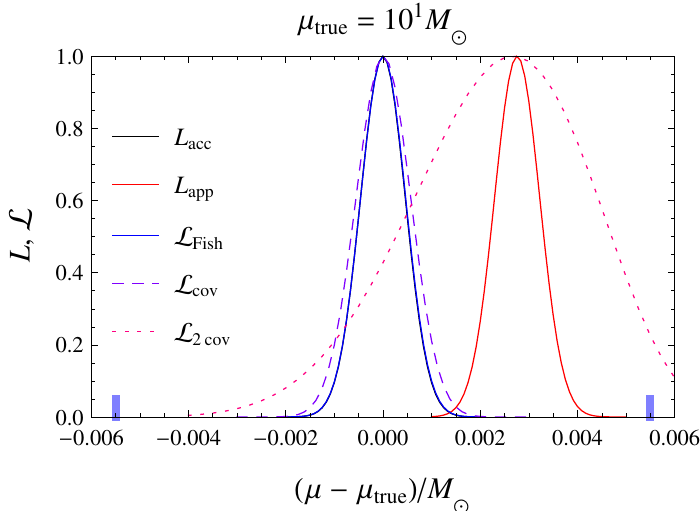}
\caption{One-dimensional likelihood plots for the standard likelihood with accurate and approximate waveforms, and the marginalized likelihood with the training sets $\mathcal{D}_\mathrm{Fish}$, $\mathcal{D}_\mathrm{cov}$ and $\mathcal{D}_\mathrm{2cov}$. The only training-set points within the horizontal plot range belong to the densest set $\mathcal{D}_\mathrm{Fish}$, and are indicated by thick marks on the horizontal axis.}
\label{fig:1D_likelihood}
\end{figure}

At a source SNR of 30, a high overlap of 0.97 between the accurate and approximate waveforms still results in a 5-sigma bias due to theoretical error; as seen in Fig.~\ref{fig:1D_likelihood}, the approximate likelihood $L_\mathrm{app}$ is peaked away from $\mu_\mathrm{true}$ with error $\mu_\epsilon\approx3\times10^{-3}M_\odot$, while the 1-sigma length for $L_\mathrm{app}$ (and the accurate likelihood $L_\mathrm{acc}$) is $\approx5\times10^{-4}M_\odot$. The marginalized likelihood with the training set $\mathcal{D}_\mathrm{Fish}$ is virtually identical to $L_\mathrm{acc}$, and with $\mathcal{D}_\mathrm{cov}$ it is slightly broader but remains peaked near the true value. For the sparsest training set $\mathcal{D}_\mathrm{2cov}$, the peak of the marginalized likelihood has an error $\mu_\epsilon$ similar to that of $L_\mathrm{app}$, although it is sufficiently broadened to ensure that it is still consistent with $\mu_\mathrm{true}$ at 2-sigma significance. These results indicate that the GPR approach can be viable for EMRIs, since even the densest considered training set $\mathcal{D}_\mathrm{Fish}$ has a grid length that is significantly larger than the width of the accurate likelihood.

For the given source and waveform parameters, a single evaluation of the NK likelihood takes 29\,s on average, as compared to an average of 5.6\,s per evaluation for the AAK likelihood. By constructing a few additional training sets with different sizes $N$, the marginalized likelihood is found to take an average of $5.6+0.01N\,\mathrm{s}$ per evaluation, i.e., $\approx2\%$ longer than the approximate likelihood for a 10-point training set. It ceases to provide computational savings over the accurate likelihood when the training set gets too large ($\gtrsim2000$ points in this particular case). However, the disparity in cost between the accurate and approximate waveforms here is only a factor of five, when in general a realistic fiducial model will be far more expensive or even completely intractable for bulk use (as seen in Sec.~\ref{subsec:LDC}).

\subsection{2-D parameter estimation}\label{subsec:2D}

\begin{figure}
\centering
\includegraphics[width=\columnwidth]{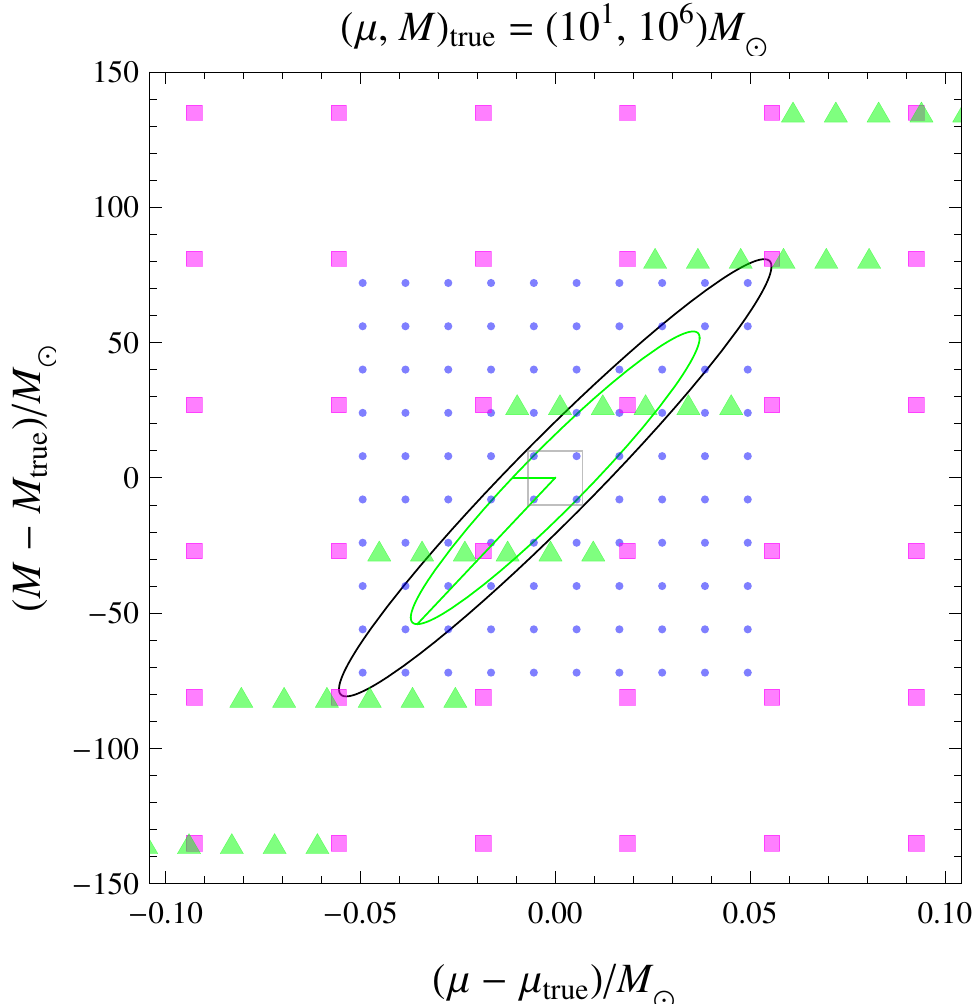}
\caption{Training-set point placement around $(\mu,M)_\mathrm{true}$ for $\mathcal{D}_\mathrm{dense}$ (dots), $\mathcal{D}_\mathrm{Fish}$ (triangles) and $\mathcal{D}_\mathrm{sparse}$ (squares). The grid for $\mathcal{D}_\mathrm{Fish}$ is defined by the semi-principal axes of the Fisher metric ellipse (green), and is aligned with the optimal covariance ellipse (black) learnt from $\mathcal{D}_\mathrm{dense}$. The central grey square corresponds to the plot range of Fig.~\ref{fig:2D_likelihood}.}
\label{fig:2D_training_set}
\end{figure}

\begin{figure}
\centering
\includegraphics[width=\columnwidth]{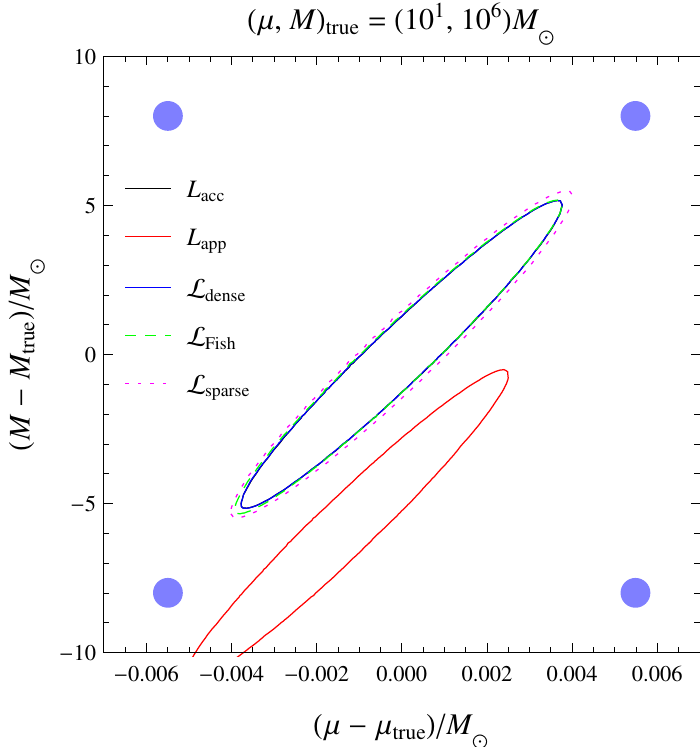}
\caption{Two-dimensional likelihood contour plots for the standard likelihood with accurate and approximate waveforms, and the marginalized likelihood with the training sets $\mathcal{D}_\mathrm{dense}$, $\mathcal{D}_\mathrm{Fish}$ and $\mathcal{D}_\mathrm{sparse}$. All contours are 2-sigma. The only training-set points within the plot range belong to $\mathcal{D}_\mathrm{dense}$, and are indicated by solid circles.}
\label{fig:2D_likelihood}
\end{figure}

In this section, the GPR marginalized likelihood \eqref{eq:marginalised_likelihood} is used to estimate the component masses $(\mu,M)_\mathrm{true}=(10^1,10^6)M_\odot$ of the source in Sec.~\ref{subsec:1D}, again assuming all other parameters are known and fixed at their true values. Maximization of the log-hyperlikelihood \eqref{eq:log-hyperlikelihood} is now over the three independent components $(g_{\mu\mu},g_{\mu M},g_{MM})$ of the covariance metric on the two-dimensional parameter subspace, with $g_{\mu\mu}g_{MM}>g_{\mu M}^2$. The eigensystem $\{(\lambda_i,\hat{\mathbf{v}}_i)\,|\,i=1,2\}$ of $\mathbf{g}$ defines a covariance ellipse with semi-principal axes $\{\lambda_i^{-1/2}\hat{\mathbf{v}}_i\}$ in the usual way.

Although the component-mass subspace is chosen for the two-dimensional example here, a straightforward search in $(\mu,M)$ is not necessarily optimal in the context of higher-dimensional parameter estimation. For example, since the central mass $M\approx\mu+M$ strongly determines the characteristic frequency of an EMRI waveform, variations in the waveform difference with respect to $M$ may be reduced by rescaling the time coordinate to dimensionless time $t/M$. This approach has been investigated, and yields longer (by an order of magnitude or so) covariance lengths as expected. However, it also results in less stable derivatives and poorer interpolation; this is likely because $M$ and $\mu+M$ are used interchangeably in the AAK model, and so its waveforms vary differently from the NK waveforms with respect to $M$. If more accurate models are used, the waveform difference will have an infinite covariance length in total mass, such that rescaling the time coordinate by $\mu+M$ reduces the component-mass subspace to a single degree of freedom (e.g., the mass ratio $\mu/M$).

Three different training sets are considered for the $(\mu,M)$ example in this section. The first is a $(6\times6)$-point set $\mathcal{D}_\mathrm{Fish}$ with $(\mu,M)_\mathrm{true}$ lying at the geometric centre of its span; its points are placed uniformly on a grid defined by the semi-principal axes $\{\lambda_i^{-1/2}\hat{\mathbf{v}}_i\}_\mathrm{Fish}$ of the Fisher metric ellipse, with $\{(\lambda_i,\hat{\mathbf{v}}_i)\}_\mathrm{Fish}$ the eigensystem of $\Gamma$ for the unit-SNR waveform difference. Two more $(10\times10)$-point sets $\mathcal{D}_\mathrm{dense}$ and $\mathcal{D}_\mathrm{sparse}$ are constructed on rectangular grids, with the grid lengths given by the short and long Fisher lengths respectively, i.e.,
\begin{equation}
(\Delta\mu,\Delta M)_\mathrm{dense}=\left(\frac{1}{\sqrt{[\Gamma]_{\mu\mu}}},\frac{1}{\sqrt{[\Gamma]_{MM}}}\right),
\end{equation}
\begin{equation}
(\Delta\mu,\Delta M)_\mathrm{sparse}=\left(\sqrt{[\Gamma^{-1}]_{\mu\mu}},\sqrt{[\Gamma^{-1}]_{MM}}\right).
\end{equation}

As justified in Sec.~\ref{subsec:1D}, the GPR model is trained on a single training set ($\mathcal{D}_\mathrm{dense}$ in this case), and the same optimal covariance ellipse is subsequently used for all three sets. The relative placement of points in the three training sets is shown in Fig.~\ref{fig:2D_training_set}, along with the covariance and Fisher ellipses. Both ellipses are aligned and the Fisher ellipse is slightly smaller, which is consistent with the discussion around \eqref{eq:covariance_lengths} and \eqref{eq:Fisher_lengths}.

From the contour plots in Fig.~\ref{fig:2D_likelihood}, the measurement of $(\mu,M)$ with the approximate likelihood $L_\mathrm{app}$ has a theoretical error of $(\mu,M)_\epsilon\approx(2\times10^{-3},6)M_\odot$, and excludes $(\mu,M)_\mathrm{true}$ at beyond 2-sigma significance. The marginalized likelihood with the training set $\mathcal{D}_\mathrm{dense}$ is virtually identical to the accurate likelihood $L_\mathrm{acc}$; so too is the likelihood for $\mathcal{D}_\mathrm{Fish}$, which is sparser and contains fewer points. More surprisingly, the training set $\mathcal{D}_\mathrm{sparse}$ also yields a likelihood that is very similar to $L_\mathrm{acc}$, which indicates that a training-set density no lower than that corresponding to the long Fisher metric lengths (i.e., the half-extents of the unit-SNR Fisher ellipse in each parameter) will still be optimal on the level of the marginalized likelihood. However, it may be difficult to learn the optimal covariance metric from such a training set if it is too sparse or contains too few points.

It is clear that a simple rectangular grid approach to the placement of training set points will not scale well with the dimensionality $\ell$ of the parameter space, but uniform placement on a grid defined by the Fisher metric eigensystem is also limited at moderately large $\ell$. From the heuristic examples studied so far, six points along each Fisher eigenvector appears to be the bare minimum for learning a covariance metric that well describes the waveform difference locally. This necessitates $O(N^3)$ operations on a $6^\ell\times6^\ell$ covariance matrix in the training stage, which is computationally challenging for $\ell>5$. However, if a suitable covariance metric can be learnt (or approximated, as in the following section), the actual set of waveform differences used in the interpolation stage does not have to be quite as large or dense as the set required to train the GPR model.

\subsection{LISA Data Challenge}\label{subsec:LDC}

The heuristic examples in Secs~\ref{subsec:1D} and \ref{subsec:2D} provide some insight into simplifying usage of the GPR marginalized likelihood \eqref{eq:marginalised_likelihood} for practical applications. Specifically, the Fisher information matrix for the unit-SNR waveform difference is closely related to the training-set grid (which is obtained through empirical validation) and the optimal covariance metric (which is obtained through training of the Gaussian process). This suggests that it could be used to fully specify both quantities at a good approximation, thus circumventing the significant computational cost associated with the two procedures.

In this section, the marginalized likelihood is applied to a quasi-realistic LISA data set containing an isolated EMRI signal. We now take the fiducial model to be the AAK, but processed through the LISACode simulator \cite{PEA2008}. The data then comprises three TDI channels $x\equiv(x_A,x_E,x_T)$ that describe the noisy response of LISA to the signal, while the inner product \eqref{eq:inner_product} generalizes to
\begin{equation}\label{eq:inner_product_TDI}
\langle a|b\rangle=4\,\mathrm{Re}\sum_{f>0}^{f_N}\mathrm{d}f\sum_{\chi=A,E,T}\frac{\tilde{a}_\chi^*(f)\tilde{b}_\chi(f)}{S_{n,\chi}(f)}
\end{equation}
in both the standard likelihood \eqref{eq:standard_likelihood} and the marginalized likelihood \eqref{eq:marginalised_likelihood}.\footnote{With the inclusion of a third independent data channel, \eqref{eq:marginalised_likelihood} also picks up an additional normalizing factor of $1/\sqrt{1+\gamma\sigma^2}$.} Waveforms that are passed through LISACode take $\gtrsim10^2\,\mathrm{s}$ to generate (on top of the original cost of the waveform), such that the standard accurate likelihood is intractable to estimate via sampling. For the approximate model, we keep the AAK as the underlying waveform, and instead apply a response that is faster and less accurate than LISACode. This is based on the FastTDI response first introduced in \cite{BGP2009}, but is adapted to the AAK model for this work; it relies on an analytic harmonic decomposition of the waveform to produce fast TDI templates directly in the frequency domain (under the stationary phase approximation).

\begin{figure}
\centering
\includegraphics[width=\columnwidth]{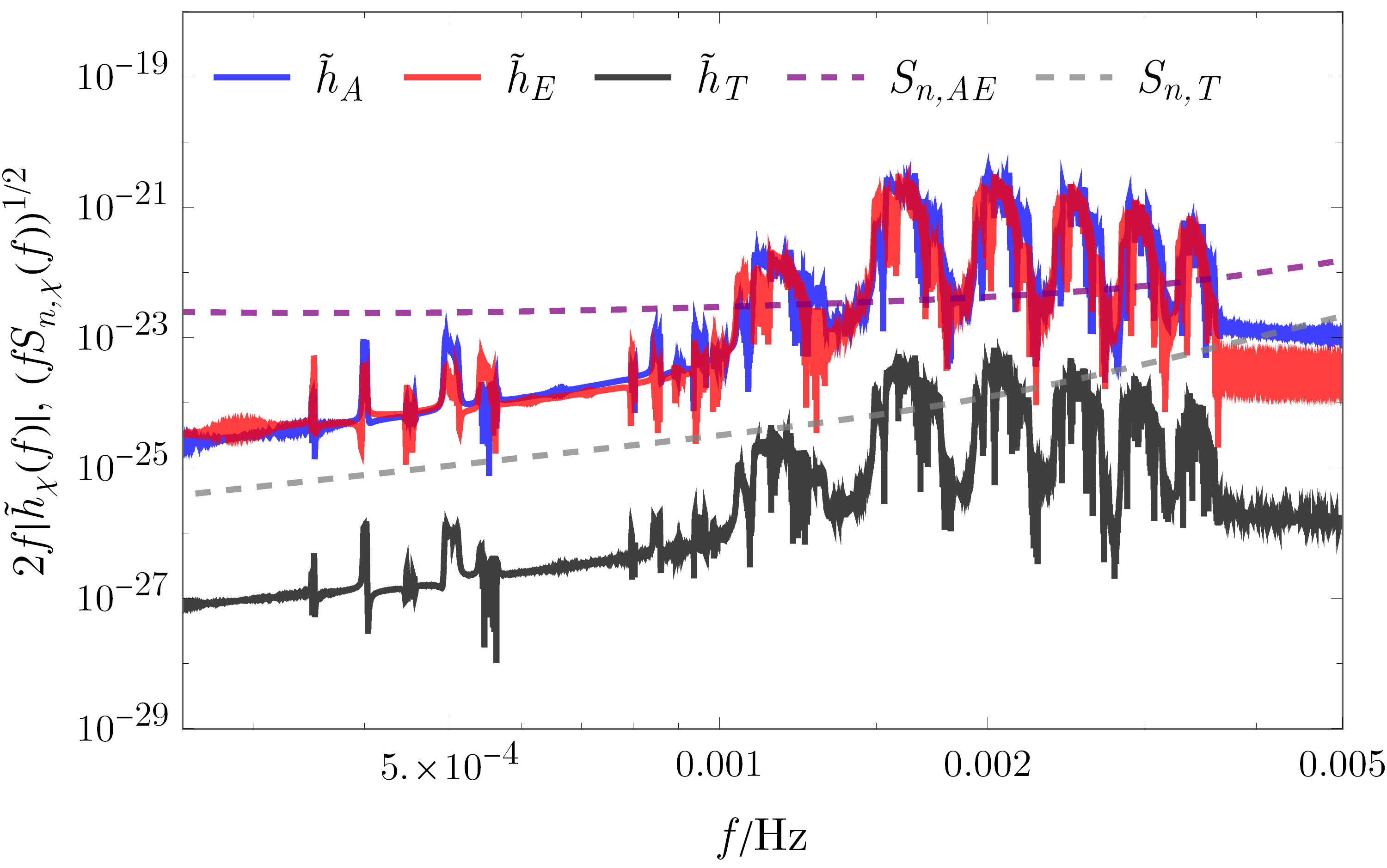}
\caption{Characteristic strain $2f|\tilde{h}_\chi(f)|$ of the injected AAK--LISACode signal, and characteristic sensitivity $(fS_{n,\chi}(f))^{1/2}$ in the three TDI channels (with $S_{n,A}=S_{n,E}$).}
\label{fig:LDC_injection}
\end{figure}

A LISACode data set that mimics the first LISA Data Challenge set \cite{LDC2019} is considered in this work (see Fig.~\ref{fig:LDC_injection}). The only differences between the two data sets are: (i) the EMRI waveform model describing the signal is taken to be the AAK instead of the older, unaugmented variant; (ii) two months of data are produced instead of two years; (iii) a different noise realization is generated, but according to the same power spectral densities used in LISACode; and (iv) the source luminosity distance is reduced from 5.2 to 0.7\,Gpc, which raises the true SNR to $\rho=29.6$ (the detection SNR for this particular noise realization is $\langle x|h\rangle/\rho=28.0$). The remaining source parameters for the signal, along with the sampling rate of 0.1\,Hz, are exactly as given in the Data Challenge set.

Even with fast templates and a relatively short two-month duration, the full parameter estimation problem remains computationally out of reach for now. We restrict the analysis to an estimation of three source parameters: the component masses $(\mu,M)$ as before, and the dimensionless spin parameter $s=a/M$. The Data Challenge parameter values are given by
\begin{equation}
(\mu,M,s)_\mathrm{true}=(29.5\,M_\odot,1.13\times10^6\,M_\odot,0.970).
\end{equation}
As it is expensive to compute numerically stable derivatives for the LISACode waveform (and hence the waveform difference), we instead use the local Fisher matrix for the unit-SNR FastTDI waveform to construct a $(6\times6\times6)$-point training set, with $(\mu,M,s)_\mathrm{true}$ lying at the geometric centre of its span; again, this is a conservative ``worst-case'' choice to ensure that the true parameters are maximally far from the nearest training-set points.

The computational cost of initializing and evaluating the marginalized likelihood is reduced by means of a low-pass filter, which is effectively applied by simply truncating both the frequency-domain data and templates at 5\,mHz (above which there is no signal information). We further streamline the analysis by foregoing the training procedure in Sec.~\ref{subsec:training}, and directly using the Fisher matrix for the approximate waveform as the metric in \eqref{eq:parameter_distance}.

The covariance scale $\sigma_f^2$ is also not treated as a hyperparameter, but is instead fixed as
\begin{equation}
\sigma_f^2=\frac{\gamma}{2N}\mathrm{tr}\,(\hat{\mathbf{K}}^{-1}\mathbf{M}),
\end{equation}
which corresponds to the analytically maximized value \eqref{eq:sigma_f}, times the empirical ratio between the waveform-difference and noise power spectral densities (averaged over frequency bins and training-set examples), i.e.,
\begin{equation}
\gamma=\frac{1}{MN}\mathrm{tr}\,[\langle h_\epsilon(\boldsymbol{\lambda}_i)|h_\epsilon(\boldsymbol{\lambda}_j)\rangle],
\end{equation}
where $M$ is the time series length ($\approx5\times10^5$ in this case). This ensures that $\gamma$ does not cancel out of the marginalized likelihood, and accounts for the fact that the average power of the waveform-difference Gaussian process is typically smaller than that of the noise; hence it prevents the estimate of statistical error from being dominated by the GPR variance, which can lead to overly stringent or even erroneous parameter estimates.

\begin{figure}
\centering
\includegraphics[width=\columnwidth]{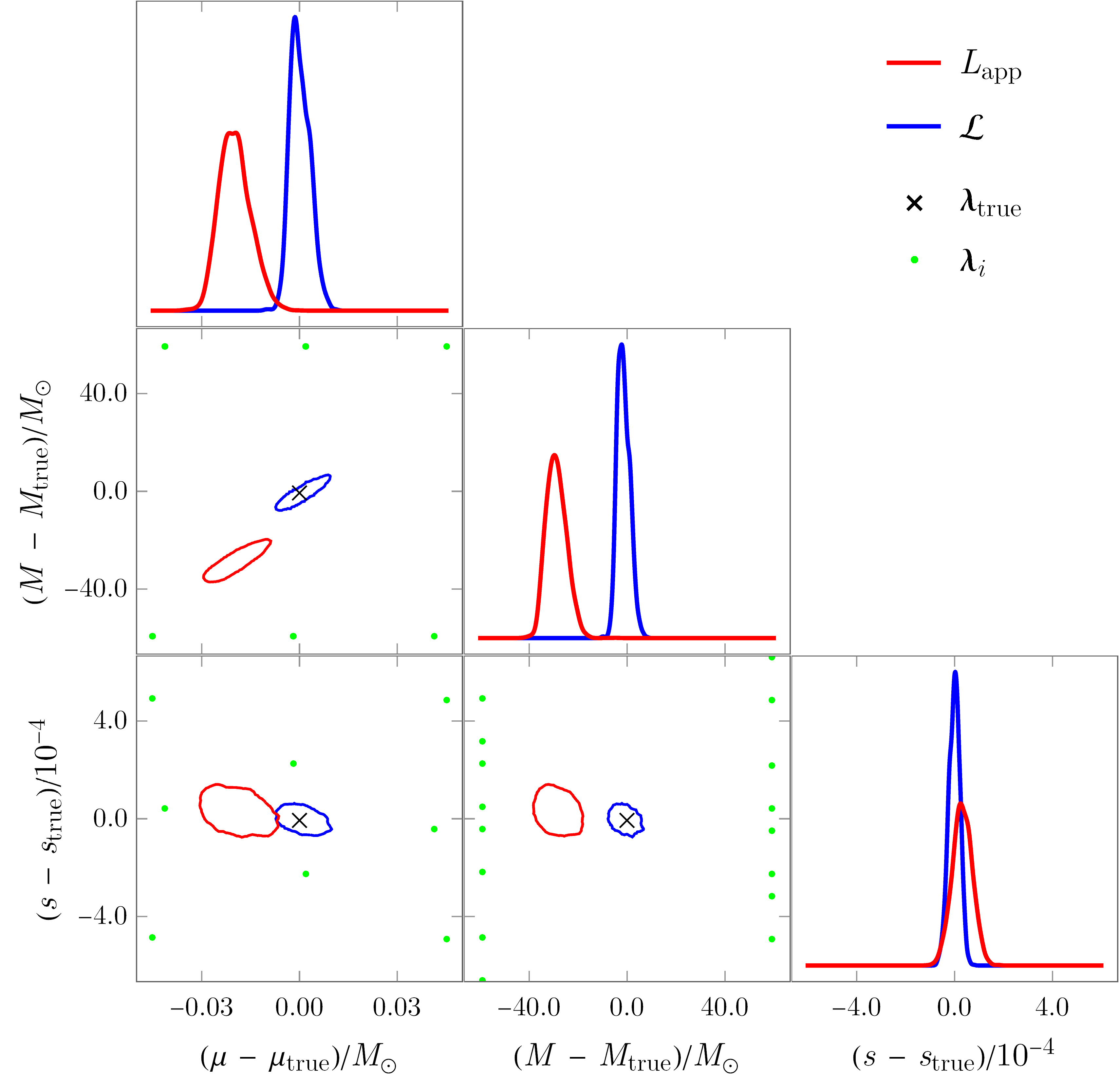}
\caption{Projected one- and two-dimensional likelihood plots for the standard likelihood with approximate waveforms, and the marginalized likelihood. All contours are 3-sigma. The true source parameters are indicated by a black cross, while the training-set points whose projections lie within the plot range are indicated by green dots.}
\label{fig:LDC_likelihood}
\end{figure}

Reconstructing the likelihood through numerical quadrature (as done in Secs~\ref{subsec:1D} and \ref{subsec:2D}) starts to become impractical in $\ell\gtrsim3$ dimensions, and so the Metropolis--Hastings algorithm \cite{H1970} is used to draw samples from both the standard approximate likelihood $L_\mathrm{app}$ and the marginalized likelihood $\mathcal{L}$. As seen in Fig.~\ref{fig:LDC_likelihood}, the approximate likelihood incurs a theoretical error of $(\mu,M,s)_\epsilon\approx(0.02M_\odot,30M_\odot,2\times10^{-5})$, which excludes the true parameters $(\mu,M,s)_\mathrm{true}$ at beyond 3-sigma significance (with $s_\mathrm{true}$ well approximated by chance). This is even though the approximate waveform itself is reasonably accurate; the overlap between the FastTDI and LISACode waveforms at $(\mu,M,s)_\mathrm{true}$, and across the span of the training set, is 0.91.

On the other hand, the marginalized likelihood remains consistent with $(\mu,M,s)_\mathrm{true}$ even in the presence of simulated LISA noise. It is also slightly more informative (precise) than the approximate likelihood, which can be attributed to the reduced fitting factor of the FastTDI waveform. The robustness of these results are verified using several different noise realizations, although only the likelihood for a single one is presented. Furthermore, the performance of the marginalized likelihood here is notwithstanding the untrained and possibly suboptimal GPR model for the waveform difference, as well as the usage of the Fisher matrix for the approximate waveform (rather than the waveform difference). This is encouraging, as such simplifications might well have to be employed in developing the method into a more extensive framework for handling theoretical error, and when integrating it within an actual EMRI analysis pipeline.

\section{Conclusion}\label{sec:conclusion}

In this paper, we have discussed the GPR marginalized-likelihood scheme \cite{MG2014,MBCG2016} in the context of EMRI data analysis, and performed a preliminary investigation of its viability for this purpose through low-dimensional studies. Even in the considered scenario where the template model used for parameter estimation has a $>90\%$ match with the source signal at the true parameter values, significant systematic bias from theoretical error will still arise for sources with moderate-to-high SNRs ($\rho\gtrsim30$). The GPR approach is shown to mitigate this bias, and hence to be suitable for improving the accuracy of EMRI parameter estimation (albeit in highly localized regions of parameter space).

The performance of the marginalized likelihood is strongly dependent on the precomputed set of waveform differences, which relies on the existence of a fiducial waveform model that reproduces the source signal with high accuracy. For the method to be practical, the density of the training set must be significantly lower than that in a notional template bank search with the fiducial waveforms. This is shown to be the case for EMRIs through a simple argument in Sec.~\ref{sec:EMRIs}, and is verified by the various examples in Secs~\ref{subsec:1D}--\ref{subsec:LDC}. Another key result in these sections is a demonstration of how the Fisher information matrix of the (normalized) waveform difference may be used to inform the placement of training-set points, as well as to estimate a covariance metric that describes the waveform difference locally.

While the marginalized likelihood shows early promise for EMRI parameter estimation, it is akin to other applications of GPR in being subject to the curse of dimensionality. The number of training-set points required to search an $\ell$-dimensional parameter subspace generally grows exponentially with $\ell$, which hinders not just the offline training stage (since the covariance matrix is larger and more ill-conditioned), but also the online interpolation stage (where a new linear combination of waveform differences is computed for each likelihood evaluation).

One possible approach to these computational problems is to replace the squared-exponential covariance function \eqref{eq:SE} in the GPR model with a covariance function that has compact support on parameter space (e.g., the Wendland polynomials \cite{W2004}), such that the covariance matrix becomes sparse. Iterative methods \cite{S2003} may then be used to accelerate the Cholesky decomposition of the covariance matrix in \eqref{eq:GPR_mean}, \eqref{eq:GPR_variance} and \eqref{eq:log-hyperlikelihood}. A compact-support covariance function also reduces the number of training-set points summed in \eqref{eq:GPR_mean}, which directly determines the evaluation speed of the marginalized likelihood.

Another strategy is to minimize the size of the training set itself. As seen throughout Sec.~\ref{sec:EMRIs}, the optimal covariance metric (or the Fisher metric) determines a fixed threshold for the characteristic density of a training set that functions well at the interpolation stage. However, it may be possible to lower this threshold density through reparametrization or dimensionality-reduction methods, e.g., the component-mass example discussed in Sec.~\ref{subsec:2D}. In the case of parameters for which this is not feasible, the number of training-set points used to cover the region of relevance may still be reduced through a non-uniform placement of points, or non-geometric prescriptions such as stochastic placement algorithms \cite{B2008,MPP2009,HAS2009}. Full coverage of the search region with a precomputed training set might not even be necessary; one possibility could be to use a ``moving'' local set that is updated adaptively as the marginalized likelihood is sampled with Markov chain Monte Carlo methods.

\begin{acknowledgments}
We thank Leor Barack, Anthony Lasenby and Christopher Berry for useful comments on the manuscript. AJKC acknowledges support from the Jet Propulsion Laboratory (JPL) Research and Technology Development program. NK is supported by a Centre National d'\'{E}tudes Spatiales fellowship. Parts of this work were carried out at JPL, California Institute of Technology, under a contract with the National Aeronautics and Space Administration. \copyright\,2020. All rights reserved.
\end{acknowledgments}

\bibliographystyle{unsrt}
\bibliography{references}
\end{document}